\documentclass[reprint,amsmath,amssymb,aps,floatfix,prl,superscriptaddress]{revtex4-1}

\usepackage{graphicx}
\usepackage{dcolumn}
\usepackage{bm}
\usepackage{epstopdf}

\begin{document}


\title{Magnetoplasmons in rotating dusty plasmas}

\author{Peter Hartmann}
\affiliation{Institute for Solid State Physics and Optics, Wigner Research Centre for Physics, Hungarian Academy of Sciences, H-1121 Budapest, Konkoly-Thege Mikl\'os str. 29-33, Hungary}
\affiliation{Center for Astrophysics, Space Physics and Engineering Research (CASPER), One Bear Place 97310, Baylor University, Waco, TX 76798, USA}

\author{Zolt\'an Donk\'o}
\affiliation{Institute for Solid State Physics and Optics, Wigner Research Centre for Physics, Hungarian Academy of Sciences, H-1121 Budapest, Konkoly-Thege Mikl\'os str. 29-33, Hungary}
\affiliation{Physics Department, Boston College, Chestnut Hill, MA 20467, USA}

\author{Torben Ott}
\author{Hanno K\"ahlert}
\author{Michael Bonitz}
\affiliation{Christian-Albrechts-Universit\"at zu Kiel, Institut f\"ur Theoretische Physik und Astrophysik, Leibnizstr. 15, 24098 Kiel, Germany}

\date{\today}

\begin{abstract}
A rotating dusty plasma apparatus was constructed to provide the possibility of experimental emulation of extremely high magnetic fields by means of the Coriolis force, observable in a co-rotating measurement frame. We present collective excitation spectra for different rotation rates with a magnetic induction equivalent of up to 3200 Tesla. We identify the onset of magnetoplasmon-equivalent mode dispersion in the rotating macroscopic two-dimensional single-layer dusty plasma. The experimental results are supported by molecular dynamics simulations of 2D magnetized Yukawa systems.
\end{abstract}


\maketitle

In the last decades, dusty plasmas have been successfully applied to study collective and transport phenomena in strongly coupled (solid and liquid) many-body systems. This is of prime interest to many fields where strong correlations are crucial, including dense astrophysical systems \cite{Peng07}, warm dense matter, trapped ions \cite{Dantan10}, ultracold neutral plasmas \cite{Killian07}, and atomic gases, see e.g. \cite{Morfill09,*BonitzRPP10}. In many of these systems magnetic fields are present (e.g. stellar plasmas) or they might be used to control the collective properties. While the interaction of magnetic fields with single particles is well understood, the mutual effect of strong correlations and magnetization remains open, and dusty plasmas are an ideal candidate to study this fundamental question.

The presence of a strong external magnetic field is expected to introduce qualitatively new features to dusty plasmas as all charge-dependent forces that act upon the dust grains can potentially be modified \cite{Ed12}. Recent numerical simulations of a strongly correlated one-component plasma model system have already made predictions regarding the collective mode structure \cite{Uchida04,Bonitz10,*Ott11,OttOCP,Farokhi12}, domain coarsening \cite{Ott13} and anomalous transport properties \cite{OttDiffu}. The direct experimental realization, although rapidly progressing, is facing serious fundamental challenges. Already in early dusty plasma experiments utilizing external magnetic fields, it was realized that the observed rotation of the dust cloud is mediated by the azimuthal component of the ion drag force and not by a direct effect of the magnetic field on the charged dust particles \cite{Konopka00,*Sato01,Schwabe11}. It is naturally the particles' charge-to-mass ratio $Q/m$ that determines the effect of external electromagnetic fields on their trajectory. Atomic plasma particles (electrons and ions) have several orders of magnitude higher $Q/m$ values than charged dust, and thus are much more sensitive to the magnetic field. In laboratory dusty plasmas, already the Earth's weak magnetic field is sufficient to slightly bend electron and ion trajectories, which drive a slow but definite rigid-body-like rotation of the dust particle ensembles in experiments \cite{Carst09}. High magnetic fields have more dramatic effects (like filamentation) on the gas discharge plasma, as shown in Refs.~\cite{Konopka05,Schwabe11} for a typical capacitively coupled radio frequency (CCRF) experimental setup, most frequently used in this type of studies. A systematic study aiming to find the operational conditions where the background plasma remains homogeneous was performed by Konopka \cite{Konopka09}. In the light of these results, we can conclude, that magnetizing the dust cloud without destroying the discharge homogeneity is extremely challenging, especially for the observation of propagating waves, as those can only be measured in a low friction (low pressure) environment, where the filamentation of the discharge is observed to be the most pronounced.

An alternative method to investigate magnetic field effects was suggested in Refs.\cite{Kahlert12,BonitzPSST}, based on the equivalence of the magnetic Lorentz force $Q({\vec v}\times{\vec B})$ and the Coriolis inertial force $2m({\vec v}\times{\vec \Omega})$ acting on moving objects when they are viewed in a reference frame rotating with frequency $\Omega$ (also called Larmor theorem). The dust particle ensemble can be forced to rotate via the drag by the background gas, driven by a rotating electrode \cite{Carst09}. Subtracting the rigid-body rotation from the trajectories of the particles makes it possible to observe the effect of the Coriolis and the centrifugal component of the inertial force, where the latter simply softens the horizontal confinement. In Ref.~\cite{Kahlert12}, experimental proof of this concept was presented by investigating the normal mode fluctuation frequency spectrum of a two-dimensional (2D = single layer) 4-particle cluster. In such a configuration, some of the frequencies are degenerate without magnetic field (rotation), which split into separate frequencies at finite magnetic fields (rotations). Due to the very different time scales characterizing the discharge plasma and the dust dynamics, we can safely neglect the effect of a few revolutions per second (rps) rotation on the background plasma, as well as on the dust charging. The magnetic -- rotating analogy is limited to vertical magnetic fields (perpendicular to the dust particle layer in laboratory experiments), but can reach equivalent magnetic induction strengths far beyond the limits of present superconducting magnets. We expect, that rotating dusty plasmas are a valuable alternative approach to overcome the fundamental difficulties of creating magnetized dusty plasmas, but only as long as the dynamics of the dust particles is concerned. Processes involving the atomic plasma constituents (e.g., dust charging, wake field formation, Coulomb shielding, plasma filamentation, etc.) are not accessible with this approach.

\begin{figure}[htb]
\includegraphics[width=1\columnwidth]{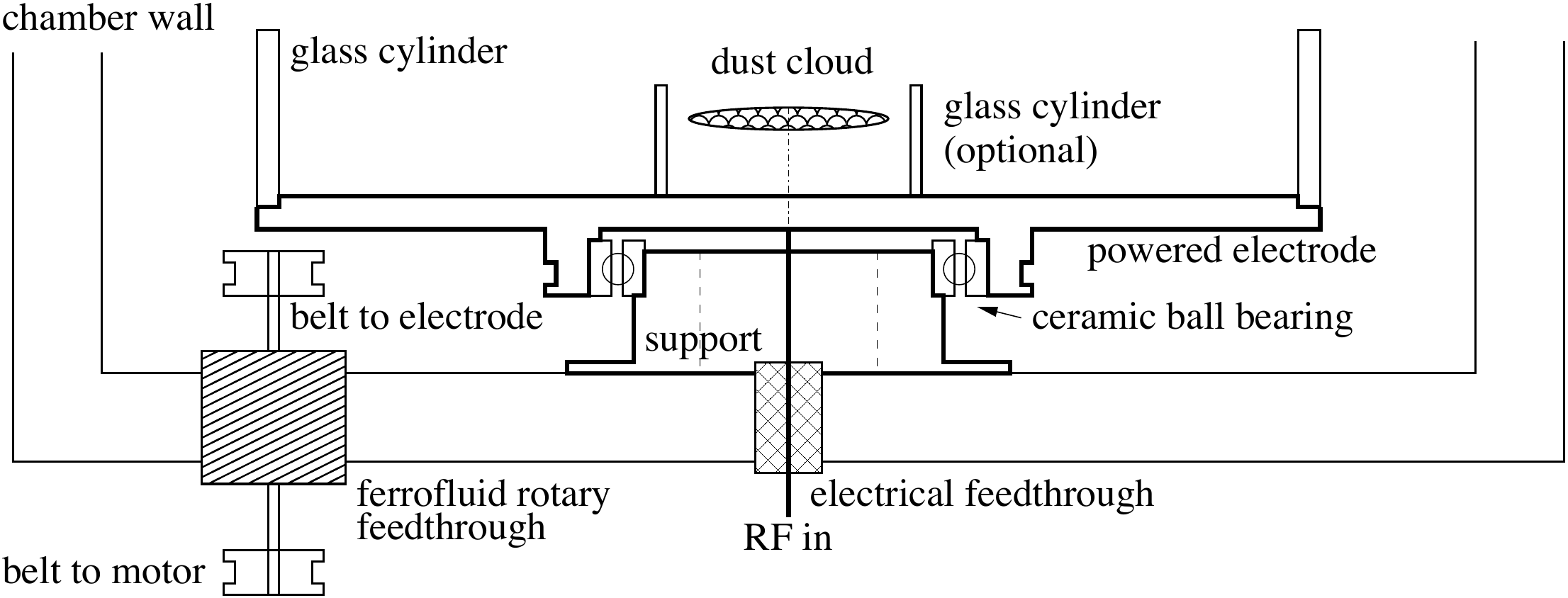}
\caption{\label{fig:setup}
Schematics of the ``RotoDust'' electrode configuration.}
\end{figure}

In this Letter we present our experimental results on wave-dispersion in 2D rotating dusty plasmas, with an equivalent magnetic induction of up to 3200 Tesla, that are expected to be relevant for a broad class of strongly correlated magnetized macroscopic systems. The experiments were performed in our dust plasma setup already introduced in Ref.~\cite{Harti11}. The lower electrode has been modified according to Fig.~\ref{fig:setup}, showing the schematic sketch of our ``RotoDust'' setup. The lower, powered electrode is mounted on a ceramic (insulating) ball bearing and features an outer glass cylinder with inner diameter of 170~mm and height of 80~mm to provide uniform gas rotation and to enhance the horizontal confinement of the dust cloud. At high rotation rates, above 1~rps, the centrifugal force becomes dominant, thus an additional small glass cylinder with inner diameter of 36~mm is placed in the center of the electrode to further enhance the electrostatic confinement. In light of this, two experimental campaigns have been performed for rotation rates 0-1.5~rps (without the small glass cylinder) and for 3-4~rps (with the small glass cylinder).

\begin{figure}[htb]
\includegraphics[width=1\columnwidth]{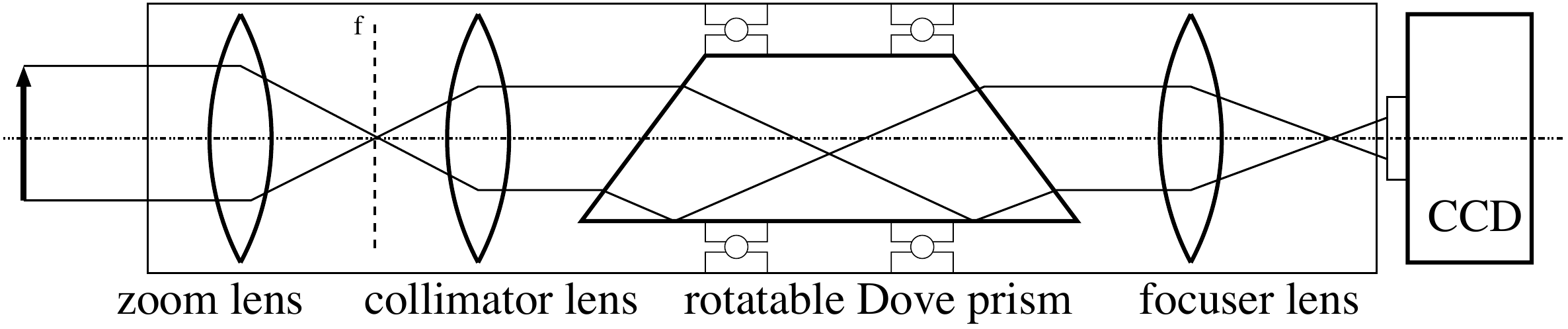}
\caption{\label{fig:optics}
Schematics of the image rotating optical configuration.}
\end{figure}

In our approach the particle trajectories have to be traced in a co-rotating frame. To realize this, we have designed an optical setup that is able to rotate the image in synchrony with the dust cloud. Our imaging system utilizes three photographic lenses and a Dove prism mounted on a rotatable support, as show schematically in Fig~\ref{fig:optics}. The video sequences were recorded with a Prosilica GX1050 camera with 100~fps at 1024x1024 pixel resolution or 200~fps at 512x512 pixel resolution for typically 5 minutes.

In the first experiment we have prepared a single layer dust cloud consisting of approximately 1000 melamine-formaldehyde (MF) particles with a diameter of 4.38~$\mu$m and mass $m=6.64\cdot10^{-14}$~kg, in a 1.0~Pa, 10~W argon RF (13.56~MHz) plasma. To obtain information on the collective excitations we use the method based on the Fourier transform of the microscopic current fluctuations, as in MD simulations \cite{Hansen75}. E.g. the longitudinal current $\lambda(k,t)= \sum_j v_{j x}(t) \exp \bigl[ i k x_j(t) \bigr]$ (utilizing the measured particle positions $x_j$ and velocities $v_j$) yields the longitudinal current fluctuations, $L(k,\omega)$ as
\begin{equation}\label{eq:L1}
L(k,\omega) = \frac{1}{2 \pi N} \lim_{\Delta T \rightarrow \infty} \frac{1}{\Delta T} | \lambda(k,\omega) |^2,
\end{equation}
where $\Delta T$ is the length of the data recording period and $\lambda(k,\omega) = {\cal{F}} \bigl[ \lambda(k,t) \bigr]$ the Fourier transform of $\lambda(k,t)$. Here, we assume that $\vec{k}$ is directed along the $x$ axis (the system is isotropic) and accordingly omit the vector notation of the wavenumber.

\begin{figure}[htb]
\includegraphics[width=1\columnwidth]{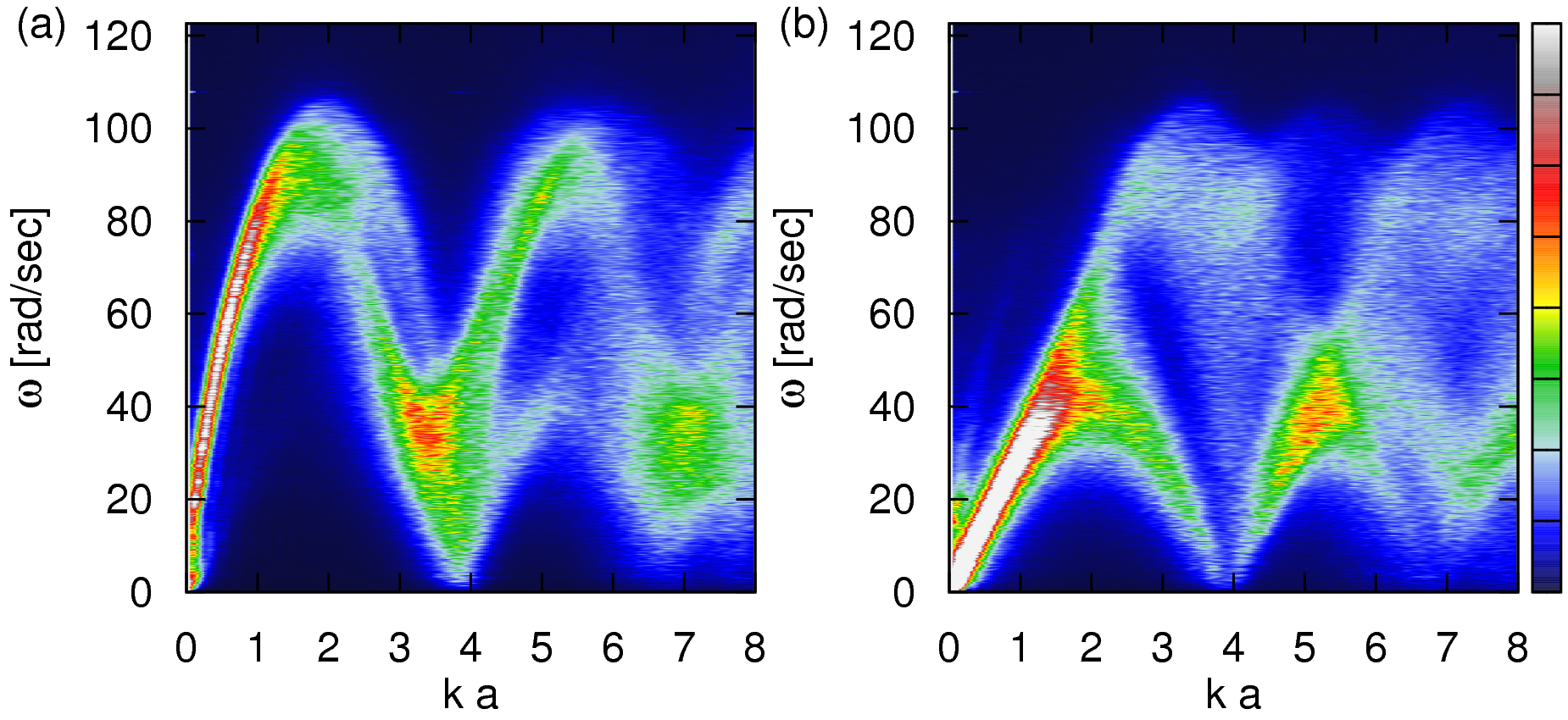}
\caption{\label{fig:rot0}
(color online) Longitudinal (a) and transverse (b) current fluctuation spectra of the single layer dust cloud without rotation ($\Omega=0$). Color scale is linear, values are in arbitrary units and are dropped.}
\end{figure}

Figure~\ref{fig:rot0} shows the longitudinal and transverse current fluctuation spectra of the system without rotation. This measurement serves as basis to determine the dust particle charge and screening length assuming Yukawa (screened Coulomb) interaction between the particles. The wavenumber is presented in a dimensionless form, normalized with the Wigner-Seitz radius $a=1/\sqrt{\pi n}$, with $n$ being the particle number density. 

The ratio of the measured longitudinal and the transverse sound speeds (slopes at $k \rightarrow 0$) is 3.7, which corresponds to a hexagonal Yukawa lattice with a screening parameter $\kappa=a/\lambda_D=0.7$ \cite{Peeters87}. For such a lattice the ratio of the nominal plasma frequency $\omega_p=Q\sqrt{n/(2\varepsilon_0 m a)}$ and the plateau frequency of the dispersion is 0.78. Matching this with our experiments resulted in the following system parameters: average dust particle charge $Q_1\approx 6600e$, Debye screening length $\lambda_{D,1}\approx 0.35$~mm. It is reasonable to expect that these parameters do not change when turning on the rotation of the electrode.

\begin{figure}[htb]
\includegraphics[width=1\columnwidth]{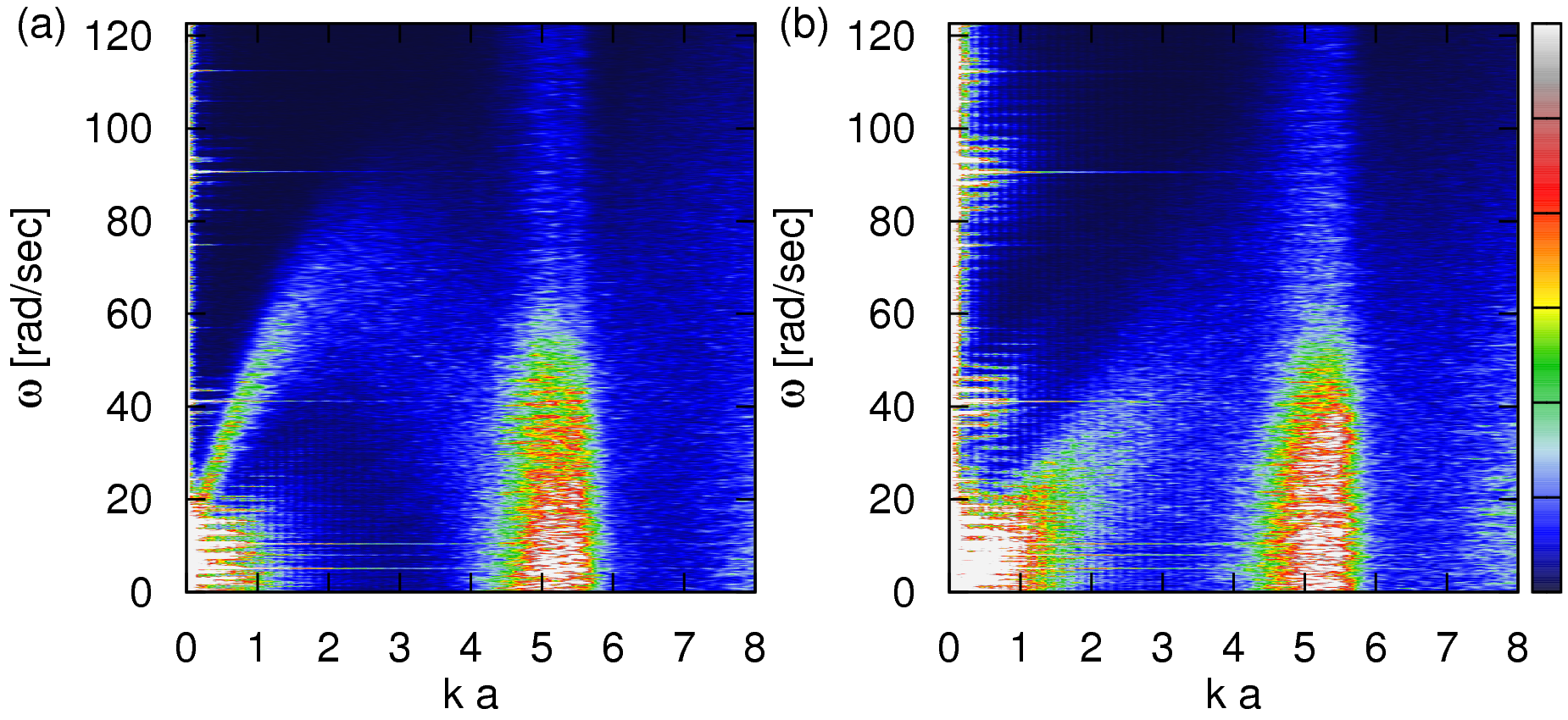}
\caption{\label{fig:rot1}
(color online) Longitudinal (a) and transverse (b) current fluctuation spectra of the single layer dust cloud with a moderate rotation of $\Omega=9.1$~rad/s.}
\end{figure}

The spectra obtained for a moderate rotation of 1.45~rps ($\Omega=9.1$~rad/s, Fig.~\ref{fig:rot1}) already show the onset of the magnetoplasmon mode in the longitudinal spectra, where the dispersion at $k=0$ of the originally acoustic mode develops a frequency gap with a magnitude of $\omega = 2\Omega = 18.2$~rad/s. The plateau frequency of the longitudinal dispersion is reduced from 100~rad/s to 85~rad/s, which, due to the reduction of the density, is an effect of the centrifugal force. Broadening of the high intensity features in the spectra indicates that -- compared to the non-rotating case -- the system is closer to the liquid state and has a lower Coulomb coupling, as predicted in Ref.~\cite{Kahlert12}. Besides these physical effects, some artifacts from the small imperfections of the experimental setup (mostly the misalignment of the optical and rotational axis) show up as multiples of the rotation frequency. 

By inserting the small glass cylinder onto the rotating electrode, we were able to increase the rotation speed while keeping the particles in the center, forming a single layer configuration. The diameter of the inner glass cylinder naturally restricts the size of the dust cloud. Due to the significantly enhanced electrostatic confinement, a large cloud that forms a stable single layer under fast rotation becomes unstable and collapses to a 3D system (sometimes called ``Yukawa ball'') when stopping the rotation. In this case, to determine the system parameters (dust charge and screening length), we had to use an alternative method to that based on the sound speeds. Here, we inserted two dust particles into the discharge plasma (0.9~Pa argon at 8~W RF power) and traced their motion without rotating the electrode. The frequency spectra of the center-of-mass and the inter-particle distance showed two distinct peaks. We assume a harmonic confinement (trap) potential in form of $V
 _\text{tr}(r) = \frac12 m \omega^2_\text{tr} r^2$ and Yukawa interaction between the particles with potential energy $V_\text{Y}(r)=Q^2\exp(-r/\lambda_D)/(4\pi\varepsilon_0 r)$. Using the measured average inter-particle distance $r_0=0.55$~mm and oscillation frequencies, the system parameters can be calculated based on the harmonic approximation solving Eqs.~(2) and (3) of Ref.~\cite{Bonitz06}:
\begin{eqnarray}
\label{eq:1}
\frac{Q^2}{4\pi\varepsilon_0} &=& \frac12 m \omega^2_\text{tr} \frac{r_0^3 \lambda_D}{r_0 + \lambda_D} e^{r_0/\lambda_D},  \\ 
\omega^2_\text{ip} &=& 3\omega^2_\text{tr}\frac{r^2_0+3r_0\lambda_D+3\lambda^2_D}{\lambda_D(r_0+\lambda_D)},   \nonumber
\end{eqnarray}
where $\omega_\text{tr}=20.1$~rad/s and $\omega_\text{ip}=44.8$~rad/s are the measured frequencies of the center-of-mass and inter-particle distance oscillations, respectively. The solution of (\ref{eq:1}) results in the system parameters of the second experimental campaign: $Q_2 = 6300e$ and $\lambda_{D,2}=0.205$~mm. Error propagation analysis shows that $\pm10$\% uncertainty in the determination of the frequencies results in ca.~$\pm17$\% variation in $Q$ and $\lambda_D$.

Collective mode fluctuation spectra were measured on ensembles of 200-300 particles. The rotation rate was individually adjusted to prepare stable, homogeneous (with variation within $\pm10$\% of the average density) single layer configurations. The systems did not show clear lattice ordering, but particle neighborhoods were stable over several periods of revolution. Based on the observed particle densities, using the previously determined charge and screening data, the parameters characterizing the investigated systems are collected in table~\ref{tab:1}.

\begin{table}[htdp]
\caption{System parameters of the experiments: $a$ -- Wigner-Seitz radius, $\Omega$ -- rotation frequency, $\kappa$ -- Yukawa parameter, $\omega_p$ -- nominal plasma frequency, $\beta$ -- magnetization parameter (ratio of the cyclotron and plasma frequencies), $B_{eq}$ -- equivalent magnetic induction.}
\begin{center}
\begin{tabular}{|l|c|c|c|}
\hline
quantity & exp. a & exp. b & exp. c \\
\hline
a [$\mu$m]       &  $326\pm30$   &  $298\pm10$    &  $398\pm10$   \\
$\Omega$ [rad/s] &  $23.3\pm0.1$ &  $22.8\pm0.1$  &  $25.0\pm0.1$ \\
$\kappa = a/\lambda_{D,2}$ & $1.59\pm0.1$ & $1.45\pm0.05$ & $1.94\pm0.1$ \\
$\omega_p$ [rad/s] & $89.2\pm3$ &  $102.1\pm4$  &  $66.1\pm2$ \\
$\beta = 2\Omega/\omega_p$ & $0.52\pm0.05$  &  $0.45\pm0.05$  &  $0.76\pm 0.08$ \\
$B_{eq} = 2m\Omega/Q$ [T] ~& ~$3065\pm150$  ~&~  $3000\pm150$  ~&~  $3290\pm160$  \\
\hline
\end{tabular}
\end{center}
\label{tab:1}
\end{table}

\begin{figure}[htb]
\includegraphics[width=1\columnwidth]{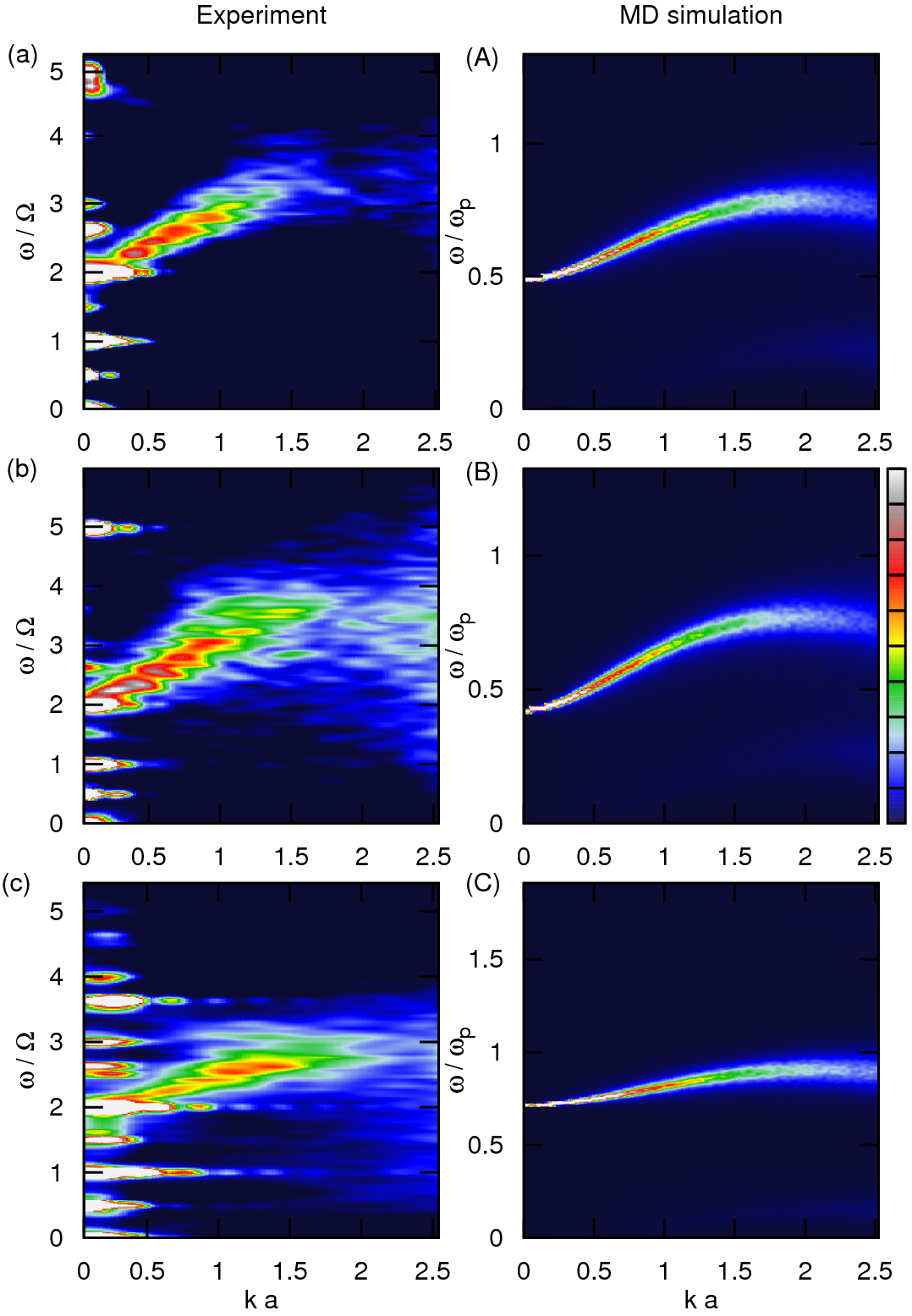}
\caption{\label{fig:rot2}
(color online) Experimental and computed longitudinal current fluctuation spectra of the single layer dust with rotation frequencies (a): $\Omega_a=23.3$~rad/s (b):  $\Omega_b=22.8$~rad/s, (c): $\Omega_c=25$~rad/s, and magnetizations (A): $\beta_a=0.52$, (B): $\beta_b=0.45$, (C): $\beta_c=0.76$, as listed in table~\ref{tab:1}. $\omega_c = \beta \omega_p$ is the cyclotron frequency.}
\end{figure}

In addition to the experiments, we have performed two-dimensional molecular dynamics (MD) simulations with periodic boundary conditions for $N=16320$ particles, which encompass a measurement time of $\omega_pt=10^5$. These simulations incorporate a magnetic field of arbitrary strength exactly through the use of a suitable integration scheme \cite{Spreiter1999} and are carried out in the microcanonical ensemble, without friction. The simulation were performed on a strongly coupled liquid system with a Coulomb coupling parameter $\Gamma=Q^2/(4\pi\varepsilon_0 a k_\text{B}T)=120$. The computed data are used to calculate the longitudinal current fluctuations according to Eq.~(\ref{eq:L1}).

Figure~\ref{fig:rot2} shows the longitudinal current fluctuation spectra for the three sets of measurements and numerical simulations of the corresponding magnetized systems. The experimental power spectra show very well developed magnetoplasmon character, with gap frequencies exactly matching the predicted value of $2\Omega$. The gap frequency in the simulations is given by the cyclotron frequency. Al low wavenumbers, the excitation frequency in both the experiment and the simulation increases with the wave number until it reaches a maximum around $ka \approx 1.5\dots 2$, depending on the parameters. At the same time, the damping of the waves increases. This behavior is in agreement with previous simulations and a theoretical analysis \cite{Bonitz10,*Ott11} based on the Quasi-Localized Charge Approximation \cite{Golden00}.

\begin{figure}[htb]
\includegraphics[width=1\columnwidth]{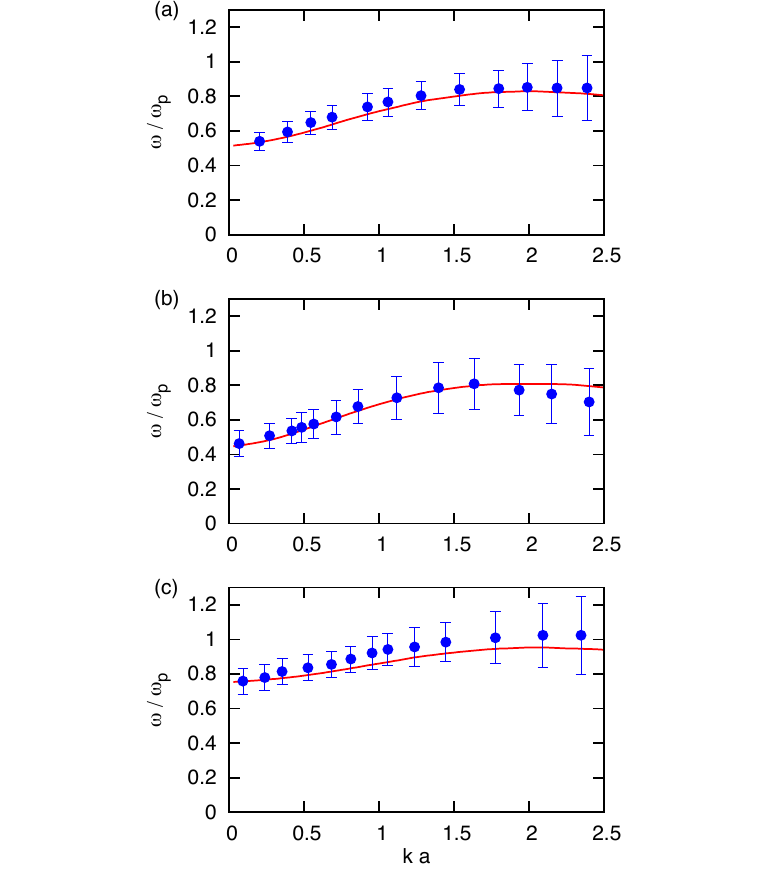}
\caption{\label{fig:rot3}
(color online) Experimental (symbols) and computed (lines) longitudinal wave dispersion. Labeling follows that of table~\ref{tab:1} and figure~\ref{fig:rot2}.}
\end{figure}

A more detailed comparison of experimental and computed wave dispersions is presented in Fig.~\ref{fig:rot3}. The best agreement is observed for $\beta=0.52$ [Fig.~\ref{fig:rot3}(a)], where even the position of the maximum is well reproduced by the simulation. In Fig.~\ref{fig:rot3}(b) [$\beta=0.47$], the agreement is excellent in the long-wavelength limit up to $ka\approx 1.5$. For smaller wavelengths the experiment shows a slightly stronger decrease of the frequency than the simulation. The agreement at the highest magnetization of $\beta=0.76$ [Fig.~\ref{fig:rot3}(c)] is still good, but the experimental spectra exhibit a higher phase velocity for small $k$, and the frequencies are consistently slightly higher than in the simulation. Considering the experimental uncertainties for the parameters entering the simulation, the agreement between the experiment and the simulation is excellent.

In summary, we have constructed a new experimental setup able to realize and observe rotating dusty plasmas in the co-rotating frame. The RotoDust setup is able to create effective magnetizations in a strongly coupled dusty plasma that are impossible to approach with superconducting magnets. At the highest rotation speed, we have achieved effective magnetic fields of $3200$~T. Converted into dimensionless parameters the effective magnetization $\beta$ reaches $0.76$ which is typical for many strongly magnetized and strongly correlated plasmas in compact astrophysical objects \cite{Kahlert12}.

The analysis of the wave spectra as observed in the rotating frame clearly shows the equivalence of the rotating dust cloud and a magnetized plasma. The $k\to 0$ gap frequency is found at twice the rotation frequency, which corresponds to the cyclotron frequency in a magnetized system. The excellent agreement with wave spectra from molecular dynamics simulations indicates the first experimental observation of the magnetoplasmon in a strongly correlated plasma. Moreover, the simulations confirm that the setup can be used for quantitative studies of extended, macroscopic strongly coupled magnetized plasmas. While we have focused on the fundamental current spectra, several other plasma properties, including diffusion~\cite{OttDiffu} or viscosity, are affected by a strong magnetic field, and are now accessible experimentally. The RotoDust setup opens the way for these investigations in the near future.

\begin{acknowledgments}
This work has been supported by the OTKA K-105476 and NN-103150 Grants, the DFG via SFB-TR24 (project A7), and a grant for CPU time at the North-German Supercomputing Alliance HLRN. 
\end{acknowledgments}


\begin{thebibliography}{27}%
\makeatletter
\providecommand \@ifxundefined [1]{%
 \@ifx{#1\undefined}
}%
\providecommand \@ifnum [1]{%
 \ifnum #1\expandafter \@firstoftwo
 \else \expandafter \@secondoftwo
 \fi
}%
\providecommand \@ifx [1]{%
 \ifx #1\expandafter \@firstoftwo
 \else \expandafter \@secondoftwo
 \fi
}%
\providecommand \natexlab [1]{#1}%
\providecommand \enquote  [1]{``#1''}%
\providecommand \bibnamefont  [1]{#1}%
\providecommand \bibfnamefont [1]{#1}%
\providecommand \citenamefont [1]{#1}%
\providecommand \href@noop [0]{\@secondoftwo}%
\providecommand \href [0]{\begingroup \@sanitize@url \@href}%
\providecommand \@href[1]{\@@startlink{#1}\@@href}%
\providecommand \@@href[1]{\endgroup#1\@@endlink}%
\providecommand \@sanitize@url [0]{\catcode `\\12\catcode `\$12\catcode
  `\&12\catcode `\#12\catcode `\^12\catcode `\_12\catcode `\%12\relax}%
\providecommand \@@startlink[1]{}%
\providecommand \@@endlink[0]{}%
\providecommand \url  [0]{\begingroup\@sanitize@url \@url }%
\providecommand \@url [1]{\endgroup\@href {#1}{\urlprefix }}%
\providecommand \urlprefix  [0]{URL }%
\providecommand \Eprint [0]{\href }%
\providecommand \doibase [0]{http://dx.doi.org/}%
\providecommand \selectlanguage [0]{\@gobble}%
\providecommand \bibinfo  [0]{\@secondoftwo}%
\providecommand \bibfield  [0]{\@secondoftwo}%
\providecommand \translation [1]{[#1]}%
\providecommand \BibitemOpen [0]{}%
\providecommand \bibitemStop [0]{}%
\providecommand \bibitemNoStop [0]{.\EOS\space}%
\providecommand \EOS [0]{\spacefactor3000\relax}%
\providecommand \BibitemShut  [1]{\csname bibitem#1\endcsname}%
\let\auto@bib@innerbib\@empty
\bibitem [{\citenamefont {Peng}\ \emph {et~al.}(2007)\citenamefont {Peng},
  \citenamefont {Brown},\ and\ \citenamefont {Truran}}]{Peng07}%
  \BibitemOpen
  \bibfield  {author} {\bibinfo {author} {\bibfnamefont {F.}~\bibnamefont
  {Peng}}, \bibinfo {author} {\bibfnamefont {E.~F.}\ \bibnamefont {Brown}}, \
  and\ \bibinfo {author} {\bibfnamefont {J.~W.}\ \bibnamefont {Truran}},\
  }\href {http://stacks.iop.org/0004-637X/654/i=2/a=1022} {\bibfield  {journal}
  {\bibinfo  {journal} {The Astrophysical Journal}\ }\textbf {\bibinfo {volume}
  {654}},\ \bibinfo {pages} {1022} (\bibinfo {year} {2007})}\BibitemShut
  {NoStop}%
\bibitem [{\citenamefont {Dantan}\ \emph {et~al.}(2010)\citenamefont {Dantan},
  \citenamefont {Marler}, \citenamefont {Albert}, \citenamefont {Gu\'enot},\
  and\ \citenamefont {Drewsen}}]{Dantan10}%
  \BibitemOpen
  \bibfield  {author} {\bibinfo {author} {\bibfnamefont {A.}~\bibnamefont
  {Dantan}}, \bibinfo {author} {\bibfnamefont {J.~P.}\ \bibnamefont {Marler}},
  \bibinfo {author} {\bibfnamefont {M.}~\bibnamefont {Albert}}, \bibinfo
  {author} {\bibfnamefont {D.}~\bibnamefont {Gu\'enot}}, \ and\ \bibinfo
  {author} {\bibfnamefont {M.}~\bibnamefont {Drewsen}},\ }\href {\doibase
  10.1103/PhysRevLett.105.103001} {\bibfield  {journal} {\bibinfo  {journal}
  {Phys. Rev. Lett.}\ }\textbf {\bibinfo {volume} {105}},\ \bibinfo {pages}
  {103001} (\bibinfo {year} {2010})}\BibitemShut {NoStop}%
\bibitem [{\citenamefont {Killian}\ \emph {et~al.}(2007)\citenamefont
  {Killian}, \citenamefont {Pattard}, \citenamefont {Pohl},\ and\ \citenamefont
  {Rost}}]{Killian07}%
  \BibitemOpen
  \bibfield  {author} {\bibinfo {author} {\bibfnamefont {T.}~\bibnamefont
  {Killian}}, \bibinfo {author} {\bibfnamefont {T.}~\bibnamefont {Pattard}},
  \bibinfo {author} {\bibfnamefont {T.}~\bibnamefont {Pohl}}, \ and\ \bibinfo
  {author} {\bibfnamefont {J.}~\bibnamefont {Rost}},\ }\href {\doibase
  http://dx.doi.org/10.1016/j.physrep.2007.04.007} {\bibfield  {journal}
  {\bibinfo  {journal} {Physics Reports}\ }\textbf {\bibinfo {volume} {449}},\
  \bibinfo {pages} {77 } (\bibinfo {year} {2007})}\BibitemShut {NoStop}%
\bibitem [{\citenamefont {Morfill}\ and\ \citenamefont
  {Ivlev}(2009)}]{Morfill09}%
  \BibitemOpen
  \bibfield  {author} {\bibinfo {author} {\bibfnamefont {G.~E.}\ \bibnamefont
  {Morfill}}\ and\ \bibinfo {author} {\bibfnamefont {A.~V.}\ \bibnamefont
  {Ivlev}},\ }\href {\doibase 10.1103/RevModPhys.81.1353} {\bibfield  {journal}
  {\bibinfo  {journal} {Rev. Mod. Phys.}\ }\textbf {\bibinfo {volume} {81}},\
  \bibinfo {pages} {1353} (\bibinfo {year} {2009})}\BibitemShut {NoStop}%
\bibitem [{\citenamefont {Bonitz}\ \emph
  {et~al.}(2010{\natexlab{a}})\citenamefont {Bonitz}, \citenamefont {Henning},\
  and\ \citenamefont {Block}}]{BonitzRPP10}%
  \BibitemOpen
  \bibfield  {author} {\bibinfo {author} {\bibfnamefont {M.}~\bibnamefont
  {Bonitz}}, \bibinfo {author} {\bibfnamefont {C.}~\bibnamefont {Henning}}, \
  and\ \bibinfo {author} {\bibfnamefont {D.}~\bibnamefont {Block}},\ }\href
  {http://stacks.iop.org/0034-4885/73/i=6/a=066501} {\bibfield  {journal}
  {\bibinfo  {journal} {Reports on Progress in Physics}\ }\textbf {\bibinfo
  {volume} {73}},\ \bibinfo {pages} {066501} (\bibinfo {year}
  {2010}{\natexlab{a}})}\BibitemShut {NoStop}%
\bibitem [{\citenamefont {{Thomas Jr.}}\ \emph {et~al.}(2012)\citenamefont
  {{Thomas Jr.}}, \citenamefont {Merlino},\ and\ \citenamefont
  {Rosenberg}}]{Ed12}%
  \BibitemOpen
  \bibfield  {author} {\bibinfo {author} {\bibfnamefont {E.}~\bibnamefont
  {{Thomas Jr.}}}, \bibinfo {author} {\bibfnamefont {R.~L.}\ \bibnamefont
  {Merlino}}, \ and\ \bibinfo {author} {\bibfnamefont {M.}~\bibnamefont
  {Rosenberg}},\ }\href {http://stacks.iop.org/0741-3335/54/i=12/a=124034}
  {\bibfield  {journal} {\bibinfo  {journal} {Plasma Physics and Controlled
  Fusion}\ }\textbf {\bibinfo {volume} {54}},\ \bibinfo {pages} {124034}
  (\bibinfo {year} {2012})}\BibitemShut {NoStop}%
\bibitem [{\citenamefont {Uchida}\ \emph {et~al.}(2004)\citenamefont {Uchida},
  \citenamefont {Konopka},\ and\ \citenamefont {Morfill}}]{Uchida04}%
  \BibitemOpen
  \bibfield  {author} {\bibinfo {author} {\bibfnamefont {G.}~\bibnamefont
  {Uchida}}, \bibinfo {author} {\bibfnamefont {U.}~\bibnamefont {Konopka}}, \
  and\ \bibinfo {author} {\bibfnamefont {G.}~\bibnamefont {Morfill}},\ }\href
  {\doibase 10.1103/PhysRevLett.93.155002} {\bibfield  {journal} {\bibinfo
  {journal} {Phys. Rev. Lett.}\ }\textbf {\bibinfo {volume} {93}},\ \bibinfo
  {pages} {155002} (\bibinfo {year} {2004})}\BibitemShut {NoStop}%
\bibitem [{\citenamefont {Bonitz}\ \emph
  {et~al.}(2010{\natexlab{b}})\citenamefont {Bonitz}, \citenamefont {Donk\'o},
  \citenamefont {Ott}, \citenamefont {K\"ahlert},\ and\ \citenamefont
  {Hartmann}}]{Bonitz10}%
  \BibitemOpen
  \bibfield  {author} {\bibinfo {author} {\bibfnamefont {M.}~\bibnamefont
  {Bonitz}}, \bibinfo {author} {\bibfnamefont {Z.}~\bibnamefont {Donk\'o}},
  \bibinfo {author} {\bibfnamefont {T.}~\bibnamefont {Ott}}, \bibinfo {author}
  {\bibfnamefont {H.}~\bibnamefont {K\"ahlert}}, \ and\ \bibinfo {author}
  {\bibfnamefont {P.}~\bibnamefont {Hartmann}},\ }\href {\doibase
  10.1103/PhysRevLett.105.055002} {\bibfield  {journal} {\bibinfo  {journal}
  {Phys. Rev. Lett.}\ }\textbf {\bibinfo {volume} {105}},\ \bibinfo {pages}
  {055002} (\bibinfo {year} {2010}{\natexlab{b}})}\BibitemShut {NoStop}%
\bibitem [{\citenamefont {Ott}\ \emph {et~al.}(2011)\citenamefont {Ott},
  \citenamefont {Bonitz}, \citenamefont {Hartmann},\ and\ \citenamefont
  {Donk\'o}}]{Ott11}%
  \BibitemOpen
  \bibfield  {author} {\bibinfo {author} {\bibfnamefont {T.}~\bibnamefont
  {Ott}}, \bibinfo {author} {\bibfnamefont {M.}~\bibnamefont {Bonitz}},
  \bibinfo {author} {\bibfnamefont {P.}~\bibnamefont {Hartmann}}, \ and\
  \bibinfo {author} {\bibfnamefont {Z.}~\bibnamefont {Donk\'o}},\ }\href
  {\doibase 10.1103/PhysRevE.83.046403} {\bibfield  {journal} {\bibinfo
  {journal} {Phys. Rev. E}\ }\textbf {\bibinfo {volume} {83}},\ \bibinfo
  {pages} {046403} (\bibinfo {year} {2011})}\BibitemShut {NoStop}%
\bibitem [{\citenamefont {Ott}\ \emph {et~al.}(2012)\citenamefont {Ott},
  \citenamefont {K\"ahlert}, \citenamefont {Reynolds},\ and\ \citenamefont
  {Bonitz}}]{OttOCP}%
  \BibitemOpen
  \bibfield  {author} {\bibinfo {author} {\bibfnamefont {T.}~\bibnamefont
  {Ott}}, \bibinfo {author} {\bibfnamefont {H.}~\bibnamefont {K\"ahlert}},
  \bibinfo {author} {\bibfnamefont {A.}~\bibnamefont {Reynolds}}, \ and\
  \bibinfo {author} {\bibfnamefont {M.}~\bibnamefont {Bonitz}},\ }\href
  {\doibase 10.1103/PhysRevLett.108.255002} {\bibfield  {journal} {\bibinfo
  {journal} {Phys. Rev. Lett.}\ }\textbf {\bibinfo {volume} {108}},\ \bibinfo
  {pages} {255002} (\bibinfo {year} {2012})}\BibitemShut {NoStop}%
\bibitem [{\citenamefont {Shahmansouri}\ and\ \citenamefont
  {Farokhi}(2012)}]{Farokhi12}%
  \BibitemOpen
  \bibfield  {author} {\bibinfo {author} {\bibfnamefont {M.}~\bibnamefont
  {Shahmansouri}}\ and\ \bibinfo {author} {\bibfnamefont {B.}~\bibnamefont
  {Farokhi}},\ }\href {\doibase 10.1017/S0022377812000116} {\bibfield
  {journal} {\bibinfo  {journal} {Journal of Plasma Physics}\ }\textbf
  {\bibinfo {volume} {78}},\ \bibinfo {pages} {259} (\bibinfo {year}
  {2012})}\BibitemShut {NoStop}%
\bibitem [{\citenamefont {Ott}\ \emph {et~al.}(2013)\citenamefont {Ott},
  \citenamefont {L\"owen},\ and\ \citenamefont {Bonitz}}]{Ott13}%
  \BibitemOpen
  \bibfield  {author} {\bibinfo {author} {\bibfnamefont {T.}~\bibnamefont
  {Ott}}, \bibinfo {author} {\bibfnamefont {H.}~\bibnamefont {L\"owen}}, \ and\
  \bibinfo {author} {\bibfnamefont {M.}~\bibnamefont {Bonitz}},\ }\href
  {\doibase 10.1103/PhysRevLett.111.065001} {\bibfield  {journal} {\bibinfo
  {journal} {Phys. Rev. Lett.}\ }\textbf {\bibinfo {volume} {111}},\ \bibinfo
  {pages} {065001} (\bibinfo {year} {2013})}\BibitemShut {NoStop}%
\bibitem [{\citenamefont {Ott}\ and\ \citenamefont {Bonitz}(2011)}]{OttDiffu}%
  \BibitemOpen
  \bibfield  {author} {\bibinfo {author} {\bibfnamefont {T.}~\bibnamefont
  {Ott}}\ and\ \bibinfo {author} {\bibfnamefont {M.}~\bibnamefont {Bonitz}},\
  }\href {\doibase 10.1103/PhysRevLett.107.135003} {\bibfield  {journal}
  {\bibinfo  {journal} {Phys. Rev. Lett.}\ }\textbf {\bibinfo {volume} {107}},\
  \bibinfo {pages} {135003} (\bibinfo {year} {2011})}\BibitemShut {NoStop}%
\bibitem [{\citenamefont {Konopka}\ \emph {et~al.}(2000)\citenamefont
  {Konopka}, \citenamefont {Samsonov}, \citenamefont {Ivlev}, \citenamefont
  {Goree}, \citenamefont {Steinberg},\ and\ \citenamefont
  {Morfill}}]{Konopka00}%
  \BibitemOpen
  \bibfield  {author} {\bibinfo {author} {\bibfnamefont {U.}~\bibnamefont
  {Konopka}}, \bibinfo {author} {\bibfnamefont {D.}~\bibnamefont {Samsonov}},
  \bibinfo {author} {\bibfnamefont {A.~V.}\ \bibnamefont {Ivlev}}, \bibinfo
  {author} {\bibfnamefont {J.}~\bibnamefont {Goree}}, \bibinfo {author}
  {\bibfnamefont {V.}~\bibnamefont {Steinberg}}, \ and\ \bibinfo {author}
  {\bibfnamefont {G.~E.}\ \bibnamefont {Morfill}},\ }\href {\doibase
  10.1103/PhysRevE.61.1890} {\bibfield  {journal} {\bibinfo  {journal} {Phys.
  Rev. E}\ }\textbf {\bibinfo {volume} {61}},\ \bibinfo {pages} {1890}
  (\bibinfo {year} {2000})}\BibitemShut {NoStop}%
\bibitem [{\citenamefont {Sato}\ \emph {et~al.}(2001)\citenamefont {Sato},
  \citenamefont {Uchida}, \citenamefont {Kaneko}, \citenamefont {Shimizu},\
  and\ \citenamefont {Iizuka}}]{Sato01}%
  \BibitemOpen
  \bibfield  {author} {\bibinfo {author} {\bibfnamefont {N.}~\bibnamefont
  {Sato}}, \bibinfo {author} {\bibfnamefont {G.}~\bibnamefont {Uchida}},
  \bibinfo {author} {\bibfnamefont {T.}~\bibnamefont {Kaneko}}, \bibinfo
  {author} {\bibfnamefont {S.}~\bibnamefont {Shimizu}}, \ and\ \bibinfo
  {author} {\bibfnamefont {S.}~\bibnamefont {Iizuka}},\ }\href {\doibase
  10.1063/1.1342229} {\bibfield  {journal} {\bibinfo  {journal} {Physics of
  Plasmas}\ }\textbf {\bibinfo {volume} {8}},\ \bibinfo {pages} {1786}
  (\bibinfo {year} {2001})}\BibitemShut {NoStop}%
\bibitem [{\citenamefont {Schwabe}\ \emph {et~al.}(2011)\citenamefont
  {Schwabe}, \citenamefont {Konopka}, \citenamefont {Bandyopadhyay},\ and\
  \citenamefont {Morfill}}]{Schwabe11}%
  \BibitemOpen
  \bibfield  {author} {\bibinfo {author} {\bibfnamefont {M.}~\bibnamefont
  {Schwabe}}, \bibinfo {author} {\bibfnamefont {U.}~\bibnamefont {Konopka}},
  \bibinfo {author} {\bibfnamefont {P.}~\bibnamefont {Bandyopadhyay}}, \ and\
  \bibinfo {author} {\bibfnamefont {G.~E.}\ \bibnamefont {Morfill}},\ }\href
  {\doibase 10.1103/PhysRevLett.106.215004} {\bibfield  {journal} {\bibinfo
  {journal} {Phys. Rev. Lett.}\ }\textbf {\bibinfo {volume} {106}},\ \bibinfo
  {pages} {215004} (\bibinfo {year} {2011})}\BibitemShut {NoStop}%
\bibitem [{\citenamefont {Carstensen}\ \emph {et~al.}(2009)\citenamefont
  {Carstensen}, \citenamefont {Greiner}, \citenamefont {Hou}, \citenamefont
  {Maurer},\ and\ \citenamefont {Piel}}]{Carst09}%
  \BibitemOpen
  \bibfield  {author} {\bibinfo {author} {\bibfnamefont {J.}~\bibnamefont
  {Carstensen}}, \bibinfo {author} {\bibfnamefont {F.}~\bibnamefont {Greiner}},
  \bibinfo {author} {\bibfnamefont {L.-J.}\ \bibnamefont {Hou}}, \bibinfo
  {author} {\bibfnamefont {H.}~\bibnamefont {Maurer}}, \ and\ \bibinfo {author}
  {\bibfnamefont {A.}~\bibnamefont {Piel}},\ }\href {\doibase
  10.1063/1.3063059} {\bibfield  {journal} {\bibinfo  {journal} {Physics of
  Plasmas}\ }\textbf {\bibinfo {volume} {16}},\ \bibinfo {eid} {013702}
  (\bibinfo {year} {2009})}\BibitemShut {NoStop}%
\bibitem [{\citenamefont {Konopka}\ \emph {et~al.}(2005)\citenamefont
  {Konopka}, \citenamefont {Schwabe}, \citenamefont {Knapek}, \citenamefont
  {Kretschmer},\ and\ \citenamefont {Morfill}}]{Konopka05}%
  \BibitemOpen
  \bibfield  {author} {\bibinfo {author} {\bibfnamefont {U.}~\bibnamefont
  {Konopka}}, \bibinfo {author} {\bibfnamefont {M.}~\bibnamefont {Schwabe}},
  \bibinfo {author} {\bibfnamefont {C.}~\bibnamefont {Knapek}}, \bibinfo
  {author} {\bibfnamefont {M.}~\bibnamefont {Kretschmer}}, \ and\ \bibinfo
  {author} {\bibfnamefont {G.~E.}\ \bibnamefont {Morfill}},\ }\href {\doibase
  10.1063/1.2134595} {\bibfield  {journal} {\bibinfo  {journal} {AIP Conference
  Proceedings}\ }\textbf {\bibinfo {volume} {799}},\ \bibinfo {pages} {181}
  (\bibinfo {year} {2005})}\BibitemShut {NoStop}%
\bibitem [{\citenamefont {Konopka}(2009)}]{Konopka09}%
  \BibitemOpen
  \bibfield  {author} {\bibinfo {author} {\bibfnamefont {U.}~\bibnamefont
  {Konopka}},\ }\href {http://psl.physics.auburn.edu/hosted/magdust} {\enquote
  {\bibinfo {title} {Complex plasma experiments with magnetic fields at MPE},}\
  } (\bibinfo {year} {2009}),\ \bibinfo {note} {workshop on Magnetized Dusty
  Plasmas, Auburn University, Auburn, Alabama, October 18 - 21}\BibitemShut
  {NoStop}%
\bibitem [{\citenamefont {K\"ahlert}\ \emph {et~al.}(2012)\citenamefont
  {K\"ahlert}, \citenamefont {Carstensen}, \citenamefont {Bonitz},
  \citenamefont {L\"owen}, \citenamefont {Greiner},\ and\ \citenamefont
  {Piel}}]{Kahlert12}%
  \BibitemOpen
  \bibfield  {author} {\bibinfo {author} {\bibfnamefont {H.}~\bibnamefont
  {K\"ahlert}}, \bibinfo {author} {\bibfnamefont {J.}~\bibnamefont
  {Carstensen}}, \bibinfo {author} {\bibfnamefont {M.}~\bibnamefont {Bonitz}},
  \bibinfo {author} {\bibfnamefont {H.}~\bibnamefont {L\"owen}}, \bibinfo
  {author} {\bibfnamefont {F.}~\bibnamefont {Greiner}}, \ and\ \bibinfo
  {author} {\bibfnamefont {A.}~\bibnamefont {Piel}},\ }\href {\doibase
  10.1103/PhysRevLett.109.155003} {\bibfield  {journal} {\bibinfo  {journal}
  {Phys. Rev. Lett.}\ }\textbf {\bibinfo {volume} {109}},\ \bibinfo {pages}
  {155003} (\bibinfo {year} {2012})}\BibitemShut {NoStop}%
\bibitem [{\citenamefont {Bonitz}\ \emph {et~al.}(2013)\citenamefont {Bonitz},
  \citenamefont {K\"ahlert}, \citenamefont {Ott},\ and\ \citenamefont
  {L\"owen}}]{BonitzPSST}%
  \BibitemOpen
  \bibfield  {author} {\bibinfo {author} {\bibfnamefont {M.}~\bibnamefont
  {Bonitz}}, \bibinfo {author} {\bibfnamefont {H.}~\bibnamefont {K\"ahlert}},
  \bibinfo {author} {\bibfnamefont {T.}~\bibnamefont {Ott}}, \ and\ \bibinfo
  {author} {\bibfnamefont {H.}~\bibnamefont {L\"owen}},\ }\href
  {http://stacks.iop.org/0963-0252/22/i=1/a=015007} {\bibfield  {journal}
  {\bibinfo  {journal} {Plasma Sources Science and Technology}\ }\textbf
  {\bibinfo {volume} {22}},\ \bibinfo {pages} {015007} (\bibinfo {year}
  {2013})}\BibitemShut {NoStop}%
\bibitem [{\citenamefont {Hartmann}\ \emph {et~al.}(2011)\citenamefont
  {Hartmann}, \citenamefont {S\'andor}, \citenamefont {Kov\'acs},\ and\
  \citenamefont {Donk\'o}}]{Harti11}%
  \BibitemOpen
  \bibfield  {author} {\bibinfo {author} {\bibfnamefont {P.}~\bibnamefont
  {Hartmann}}, \bibinfo {author} {\bibfnamefont {M.~C.}\ \bibnamefont
  {S\'andor}}, \bibinfo {author} {\bibfnamefont {A.}~\bibnamefont {Kov\'acs}},
  \ and\ \bibinfo {author} {\bibfnamefont {Z.}~\bibnamefont {Donk\'o}},\ }\href
  {\doibase 10.1103/PhysRevE.84.016404} {\bibfield  {journal} {\bibinfo
  {journal} {Phys. Rev. E}\ }\textbf {\bibinfo {volume} {84}},\ \bibinfo
  {pages} {016404} (\bibinfo {year} {2011})}\BibitemShut {NoStop}%
\bibitem [{\citenamefont {Hansen}\ \emph {et~al.}(1975)\citenamefont {Hansen},
  \citenamefont {McDonald},\ and\ \citenamefont {Pollock}}]{Hansen75}%
  \BibitemOpen
  \bibfield  {author} {\bibinfo {author} {\bibfnamefont {J.~P.}\ \bibnamefont
  {Hansen}}, \bibinfo {author} {\bibfnamefont {I.~R.}\ \bibnamefont
  {McDonald}}, \ and\ \bibinfo {author} {\bibfnamefont {E.~L.}\ \bibnamefont
  {Pollock}},\ }\href {\doibase 10.1103/PhysRevA.11.1025} {\bibfield  {journal}
  {\bibinfo  {journal} {Phys. Rev. A}\ }\textbf {\bibinfo {volume} {11}},\
  \bibinfo {pages} {1025} (\bibinfo {year} {1975})}\BibitemShut {NoStop}%
\bibitem [{\citenamefont {Peeters}\ and\ \citenamefont {Wu}(1987)}]{Peeters87}%
  \BibitemOpen
  \bibfield  {author} {\bibinfo {author} {\bibfnamefont {F.~M.}\ \bibnamefont
  {Peeters}}\ and\ \bibinfo {author} {\bibfnamefont {X.}~\bibnamefont {Wu}},\
  }\href {\doibase 10.1103/PhysRevA.35.3109} {\bibfield  {journal} {\bibinfo
  {journal} {Phys. Rev. A}\ }\textbf {\bibinfo {volume} {35}},\ \bibinfo
  {pages} {3109} (\bibinfo {year} {1987})}\BibitemShut {NoStop}%
\bibitem [{\citenamefont {Bonitz}\ \emph {et~al.}(2006)\citenamefont {Bonitz},
  \citenamefont {Block}, \citenamefont {Arp}, \citenamefont {Golubnychiy},
  \citenamefont {Baumgartner}, \citenamefont {Ludwig}, \citenamefont {Piel},\
  and\ \citenamefont {Filinov}}]{Bonitz06}%
  \BibitemOpen
  \bibfield  {author} {\bibinfo {author} {\bibfnamefont {M.}~\bibnamefont
  {Bonitz}}, \bibinfo {author} {\bibfnamefont {D.}~\bibnamefont {Block}},
  \bibinfo {author} {\bibfnamefont {O.}~\bibnamefont {Arp}}, \bibinfo {author}
  {\bibfnamefont {V.}~\bibnamefont {Golubnychiy}}, \bibinfo {author}
  {\bibfnamefont {H.}~\bibnamefont {Baumgartner}}, \bibinfo {author}
  {\bibfnamefont {P.}~\bibnamefont {Ludwig}}, \bibinfo {author} {\bibfnamefont
  {A.}~\bibnamefont {Piel}}, \ and\ \bibinfo {author} {\bibfnamefont
  {A.}~\bibnamefont {Filinov}},\ }\href {\doibase
  10.1103/PhysRevLett.96.075001} {\bibfield  {journal} {\bibinfo  {journal}
  {Phys. Rev. Lett.}\ }\textbf {\bibinfo {volume} {96}},\ \bibinfo {pages}
  {075001} (\bibinfo {year} {2006})}\BibitemShut {NoStop}%
\bibitem [{\citenamefont {Spreiter}\ and\ \citenamefont
  {Walter}(1999)}]{Spreiter1999}%
  \BibitemOpen
  \bibfield  {author} {\bibinfo {author} {\bibfnamefont {Q.}~\bibnamefont
  {Spreiter}}\ and\ \bibinfo {author} {\bibfnamefont {M.}~\bibnamefont
  {Walter}},\ }\href {\doibase DOI: 10.1006/jcph.1999.6237} {\bibfield
  {journal} {\bibinfo  {journal} {J. Comput. Phys.}\ }\textbf {\bibinfo
  {volume} {152}},\ \bibinfo {pages} {102 } (\bibinfo {year}
  {1999})}\BibitemShut {NoStop}%
\bibitem [{\citenamefont {Golden}\ and\ \citenamefont
  {Kalman}(2000)}]{Golden00}%
  \BibitemOpen
  \bibfield  {author} {\bibinfo {author} {\bibfnamefont {K.~I.}\ \bibnamefont
  {Golden}}\ and\ \bibinfo {author} {\bibfnamefont {G.~J.}\ \bibnamefont
  {Kalman}},\ }\href {\doibase 10.1063/1.873814} {\bibfield  {journal}
  {\bibinfo  {journal} {Physics of Plasmas}\ }\textbf {\bibinfo {volume} {7}},\
  \bibinfo {pages} {14} (\bibinfo {year} {2000})}\BibitemShut {NoStop}%
\end{thebibliography}

%

\end{document}